\pdfoutput=1

\documentclass[prl,10pt,twocolumn,superscriptaddress,showpacs]{revtex4}
\usepackage{amsmath}
\usepackage{latexsym}
\usepackage{amssymb}
\usepackage{graphics,epstopdf}
\usepackage{graphicx}
\usepackage[colorlinks=true, citecolor=blue, urlcolor=blue ]{hyperref}
\usepackage{float}
\usepackage{graphicx}
\usepackage{amsfonts}
\usepackage{soul}

\begin{document}
\title{Probing hierarchy of temporal correlation requires either generalised measurement or nonunitary evolution}

\author{Shiladitya Mal}
\email{shiladitya.27@gmail.com}
\affiliation{S. N. Bose National Centre for Basic Sciences, Salt Lake, Kolkata 700 098, India.}

\author{A. S. Majumdar}
\email{archan@bose.res.in}
\affiliation{S. N. Bose National Centre for Basic Sciences, Salt Lake, Kolkata 700 098, India.}

\author{D. Home}
\email{qunatumhome@gmail.com}
\affiliation{Center for Astroparticle Physics and Space Science , Bose Institute, Kolkata 700091, India}

\begin{abstract}
Temporal steering and violation of the Leggett-Garg inequality are two different ways of probing 
the violation of macro-realistic assumptions in quantum mechanics. It is shown here that under 
unitary evolution and projective measurements the two types of temporal correlations lead to 
similar results. However, their inequivalence is revealed if either one of them is relaxed, i.e., by employing either generalized measurements, or noisy evolution, as we show here using relevant examples.
\end{abstract}


\maketitle

\section{I. Introduction}

There exist several counter intuitive phenomena in quantum theory without parallel in 
classical mechanics,  going beyond mere theoretical curiosity to the realm of 
useful practical applications.  
Different types of entanglement have been used as resources in  tasks such as secure key 
generation~\citep{ekrt}, random number generation~\citep{rand}, and distinguishing between 
quantum channels~\citep{pian}. Bell-nonlocality and Einstein-Podolsky-Rosen (EPR) 
steering are two types of quantum correlations that may be present in spatially 
correlated entangled systems. Violation of Bell-CHSH inequalities~\cite{bnl,chsh} prohibits the
local hidden variable description of correlations in measurement outcomes corresponding to the
two entangled parties.  Steering, originally introduced by Schr¨odinger~\citep{schro} in response 
to the EPR paradox~\citep{epr}, is the ability of one party, Alice, to affect the state of 
another remote party, Bob, through her choice of measurements. Such an ability depends
 on the entanglement of the pair shared between Alice and Bob, as well as the measurement 
settings chosen for each particles of the pair. 

The concepts of Bell-nonlocality and EPR steering have both been reformulated in terms of 
information theoretic tasks~\citep{eprt} leading to characterization of entanglement between 
two parties under various levels of trust assigned to the measurement devices possessed by them.
 Such a characterization has applications
in the secret key generation protocols, {\it i.e.,} security through violation of Bell-inequality
in device independent quantum key distribution(QKD)~\citep{diqkd}, and through steering in  
one-sided device-independent QKD~\citep{1diqkd}. The hierarchy 
of spatial correlations
in quantum mechanics is thereby revealed, {\it i.e.}, the set of all states 
violating Bell-CHSH inequalities form
a strict subset of the the set of all states that are steerable, which in turn themselves form 
a strict subset of the set of all entangled states~\citep{eprt}.  

The subject of temporal correlations in quantum mechanics has attracted increased attention in
recent years. Temporal correlations refer to the correlations in the outcomes of measurements
performed on the same particle acquired through two or more successive measurements, as distinct
from spatial correlations which pertain to correlations in outcomes of measurements
performed on two or more spatially separated particles. Following the seminal work of Leggett 
and Garg~\citep{lgi}, various studies
have been performed leading to the formulation of macrorealistic inequalities
for temporally correlated systems. The Leggett-Garg inequalities (LGI) have been generalized
for  different physical systems~\citep{lgplus}. A bound for temporal correlation has been derived~\citep{bdrni} by considering a general scenario whose special cases arise in the context of macrorealism and noncontextuality. The issue of how classicality may emerge for
temporally correlated systems have also been studied~\citep{class,kb,mal,sbm}.
It has been further shown~\citep{ck} that necessary and sufficient conditions for macro-realism 
emerge when no signalling in time (NSIT) is satisfied for all combinations of sequential measurements.

Several experimental results 
on violations of LGI have been reported\citep{elgi}, signifying the untenability of macrorealistic
principles at the quantum level, parallel to the refutation of local realism through experimental 
violations of Bell-CHSH inequalities. On the other hand, the notion of steering has been
extended recently in the domain of temporal correlations through the formulation of
single system steering where measurements on a single system are considered at different
times~\citep{tepster}. The obtained temporal steering inequality is related to the security
bound of the Bennett-Brassard 1984~\citep{bb84} protocol of QKD. It has been shown~\citep{tepster}
that the temporal steering inequality is formally mathematically equivalent to an inequality
for spatial steering. 
 
Quantum invasiveness of measurement is ingrained in the above two different formulations of 
temporal correlations probed through the LGI and the temporal steering inequality. There are three alternative  conditions for probing the violation of macrorealism for temporal correlations, {\it viz.}, through the violation of LGI, NSIT, or WLGI (Wigner type LGI). A comparative study of these alternative conditions has been done in Ref.~\citep{mal}.
Violation of LGI implies there is no underlying hidden variable model(HVM) reproducing all the temporal correlations. Temporal steering on the other hand, is probed through the violation of a weaker notion of the non-existence of a 
hidden state model, i.e., hidden state model (HSM) for the post-measurement state.  Though a  hidden variable model (HVM) trivially follows from any HSM, it is nontrivial to show how these different notions of temporal correlations differ in real testable situations. There exists no study in the literature along these lines. 

Therefore, a natural question arises here as to what, if any, is
the relation between these two apparently different ways of probing quantum invasiveness.
The motivation for the present work is to probe the similarities and differences pertaining to temporal
correlations formulated in the above two ways. To this end in the present
work we first describe our protocol of temporal steering and derive a temporal analogue of the CHSH type steering inequality~\cite{bv}.  We then show that steerable and LGI violating correlations are equivalent in the context of unitary time evolution and projective measurements.

The degeneracy between the two types of temporal correlations
pertaining to the violation of LGI on the one hand, and single system steering on the other, is
thus evident at the level of unitary evolution and standard projective measurements. In order to
break this degeneracy, we next consider more general measurements. The hierarchy between the two
different kinds of temporal correlations is revealed  by employing generalised quantum measurements dubbed as positive operator 
valued measurements (POVM). 
The dissimilarity with the case of spatial correlation is apparent since only projective measurements are sufficient for differentiating Bell nonlocality and spatial steerability.  We finally consider the case of non-unitary evolution with 
projective measurements, which demonstrates that the LGI violating correlations are stronger 
than those that allow temporal steering. 

\section{II. Different kinds of temporal correlations}

We begin by describing  the scenario of temporal correlations of a single system undergoing
time evolution. Temporal correlations have been
 considered in a general form in Ref~.\citep{bdrni} irrespective of the type of evolution or compatibility of the observables sequentially measured. Measurement of commutative observables sequentially on single system invokes the scenario of quantum contextuality test, whereas noncommutative observables correspond to a test of macrorealism. Here we consider different possible types of temporal correlation in the context of noncommutative sequential measurements.
 
{\it Protocol of single system steering:} Sequential measurements on a single system is usually done by a single observer measuring at different times. It can also be thought as if there are two observers, say, Alice and Bob measuring on a single system sequentially. In our steering protocol in a single run Alice measures on the system first according to Bob's request, and then sends it to him via some quantum channel, announcing her result publicly. Finally, Bob measures on the system in his possession.
A system evolves from an initial state $\rho(t=0)$ to some state 
$\rho(t)$ at time $t$ under some quantum channel. After performing a general
measurement (POVM) on the system as asked by Bob, Alice sends it to him via a quantum channel.
 For example, Alice can create an assemblage 
like $\lbrace\tilde{\rho}(a_{k}|k)\rbrace$ according to her choice of measurement  
$\lbrace M^{k}(a_{k})\rbrace$. Unnormalised states are denoted by tilde on $\rho$.  Here $a_{k}$ denotes the outcome $a$ of the $k$-th POVM and 
$k\in\lbrace 1,2...n\rbrace$;  $a\in\lbrace 0,1...d\rbrace$. The set of POVMs satisfies 
$M^{k}(a_{k})\geq 0$ and $\sum_{a}M^{k}(a_{k})=\mathbb I$. The probability of getting $a_{k}$ is given by 
$p(a_{k})=tr[\rho M^{k}(a_{k})]$. 

\begin{figure}[t!]
\centering
\includegraphics[height=5cm,width=7cm]{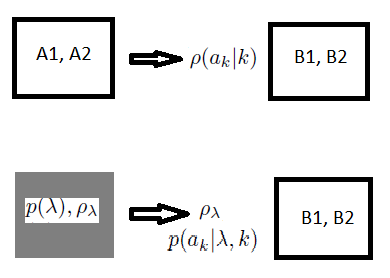}
\caption{ Schematic diagram for single system steering when each party measures from different sets of two observables. (i) Alice measures $A_{1}$ or $A_{2}$ according to the request from Bob, sends the post-measurement state to him through some quantum channel and announces her outcome publicly. (ii) Classical mimicry of the above case. Alice picks $\rho_{\lambda}$ with probability $p_{\lambda}$, announces outcome $a_{k}$ with probability $p(a_{k}|\lambda, k)$ and sends the particle to Bob. }\label{scheme}
\end{figure} 

Bob needs to verify whether Alice gave him a post-measured assemblage 
$\lbrace \tilde{\rho}(a_{k}|k)=\sqrt{M^{k}(a_{k})}\rho\sqrt{M^{k}(a_{k})}\rbrace$ or not.  If Bob finds after 
state tomography that the assemblage $\lbrace\tilde{\rho}(a_{k}|k)\rbrace$ has come from a fixed ensemble 
$\lbrace p(\lambda), \rho_{\lambda}\rbrace$, called hidden state ensemble (HSE), (where $p(\lambda)$ is 
the distribution of states $\rho_{\lambda}$),  he is not convinced about Alice's steerability. In this 
case Alice can adopt a classical strategy (cheating strategy) to prepare the state for Bob, which can be written in terms of HSE as
\begin{eqnarray}\label{hs}
\tilde{\rho}(a_{k}|k)=\sum_{\lambda}p(\lambda) p(a_{k}|\lambda,k)\rho_{\lambda}
\end{eqnarray} 
This means Alice has sent Bob states by picking from the
HSE with associated weights and declares outcome `$a_k$' with probability $p(a_{k}|\lambda,k)$, where $0\leq p(a_{k}|\lambda,k)\leq 1, \sum_{a}p(a_{k}|\lambda,k)=1$.

Now let us see how joint probabilities for sequential measurements can be constructed according to the classical strategies and indicate some quantum information processing tasks outperforming such classical strategies. Suppose Alice measures $A_{1}$ or $A_{2}$ at an earlier time $t_A$ on the system, 
and Bob measures $B_{1}$ or $B_{2}$ at a later time $t_B$. Here, three types of possibilities can arise:\\

(i) In the \emph{ non-invasive realist model} (NIRM), a hidden variable model (HVM) pertinent to the LG scenario, the joint probabilities can be written as
\begin{eqnarray}\label{nirm}
P(A_{i}=a_{i}, B_{j}=b_{j})=\sum_{\lambda}p(\lambda)p(a_{i}|A_{i}, \lambda) p(b_{j}|B_{j}, \lambda)
\end{eqnarray}
This means Alice and Bob obtain their outcome $a_{i},b_{j}$ when measuring $A_{i},B_{j}$ according to some predetermined strategy $\lambda$. This NIRM (\ref{nirm}) leads to an LGI~\citep{sa}. 
Quantum violation of this inequality has been linked with information processing tasks such as saving memory in computing~\citep{bv}, in the context of QKD~\citep{lgqkd} and randomness generation~\citep{mms}. As only classical communication between two parties can simulate violation of LGI, it should be emphasised that fully device independent information processing tasks can not be devised with temporal correlations without any other restrictions. \\

  (ii) There exists a hidden state model (HSM) for Bob when Alice is not capable of steering, and joint probabilities can be written as
\begin{eqnarray}\label{ns}
P(A_{i}=a_{i}, B_{j}=b_{j})=\sum_{\lambda}p(\lambda)p(a_{i}|A_{i}, \lambda) p^{Q}(b_{j}|B_{j},\rho_{\lambda})
\end{eqnarray}
Violation of any inequality derived from Eq.(\ref{ns}) is a
demonstration of temporal steering. When this is valid for all measurements performed by Alice and Bob, Eq.(\ref{ns}) actually coincides with Eq.(\ref{hs}). Later in this paper we shall use the words HSM and HSE in the same sense. Quantum violation of TSI enables  secure key generation under  coherent attack by cloning~\citep{stwit}. Some semi-device independent tasks can be formulated in this case but we do not focus on this issue here.\\

(iii) In the case when it is known that Alice sends quantum systems for Bob's measurement, they can always generate a secure key relying on 
the coherence property of the state and the uncertainty relation according to the original BB84 
protocol~\citep{bb84}. It may be mentioned here that entanglement in space is described by the tensor 
product structure between states belonging to  different Hilbert spaces corresponding to spatially 
separated systems. However, time evolution leads to the transformation of states in the same Hilbert 
space. Hence, it is not possible to describe states separated in time by a similar tensor product 
structure, and correspondingly, there exists no analogous separability criterion for temporal
correlations.

\section{III. Equivalence under unitary evolution and projective measurements}

In spite of the obvious differences in the joint probability distributions given by Eqs.(\ref{nirm}) 
and (\ref{ns}) for the two distinct cases (i) and (ii), respectively,
we now show that they are equivalent under unitary channel and projective measurements. To prove this we introduce a lemma of optimal ensemble for the temporal scenario adapted from the results derived in the context of spatial correlations~\citep{eprt}. Existence of optimal ensemble ensures that there cannot be any other ensemble which satisfies Eq.(\ref{hs}) for temporal correlations, iff the optimal ensemble cannot satisfy it.\\

{\bf Lemma 1}: Consider a group G with unitary representation $U(g)$ on the Hilbert space of the system. Suppose, $\forall A\in\mathcal{M}_{A}$ (which Alice can measure), $\forall a$ and $\forall g\in G $, if we have $U_{1}^{\dagger}(g)AU_{1}(g)\in\mathcal{M}_{A}$ and $\tilde{\rho}_{a}^{U_{1}^{\dagger}(g)AU_{1}(g)}=U_{2}(g)\tilde{\rho}_{a}^{A}U_{2}^{\dagger}(g)$, then there exists a G-covariant optimal ensemble: $\lbrace \rho_{\lambda}^{\star}, p_{\lambda}^{\star}\rbrace = \lbrace U_{2}(g)\rho_{\lambda}^{\star}U_{2}^{\dagger}(g), p_{\lambda}^{\star}\rbrace$. $U_{1}(g)$ and $U_{2}(g)$ are unitary operations applied by Alice and Bob respectively.\\

{\bf Proof}: Suppose a HSE $\lbrace p_{\lambda}, \rho_{\lambda}\rbrace$ exists such
that $\tilde{\rho}_{a}^{A}=\sum p_{\lambda}\rho_{\lambda}p(a|A,\lambda)$. Then we have $U_{2}(g)\tilde{\rho}_{a}^{A}U_{2}^{\dagger}(g)=\sum p_{\lambda}^{\star}U_{2}(g)\rho_{\lambda}U_{2}^{\dagger}(g)p(a|A,\lambda)$ and $\tilde{\rho}_{a}^{U_{1}^{\dagger}(g)AU_{1}(g)}=\sum p_{\lambda}\rho_{\lambda}p(a|U_{1}^{\dagger}(g)AU_{1}(g))$. Now applying the conditions of the lemma 1, we can derive the G-covariant optimal ensemble $F^{\star}=\lbrace p_{\lambda}^{\star}, U_{2}(g)\rho_{\lambda}U_{2}^{\dagger}(g)\rbrace$ with $p_{\lambda}^{\star}=p_{\lambda}d\mu_{G(g)}$ with the choice $p(a|A,\lambda)=p(a|U_{1}^{\dagger}(g)A U_{1}(g),\lambda)$. \\

Now we are in a position to prove a theorem. For any initial state under unitary evolution and projective measurements lemma 1 holds. This is because firstly $U_{1}^{\dagger}(g)AU_{1}(g)\in\mathcal{M}_{A}$. For the other condition,  i.e., $\tilde{\rho}_{a}^{U_{1}^{\dagger}(g)AU_{1}(g)}=U_{2}(g)\tilde{\rho}_{a}^{A}U_{2}^{\dagger}(g)$, suppose, Alice measures $U_{1}^{\dagger}(g)AU_{1}(g)$, then the unnormalised state becomes $\tilde{\rho}_{a}^{U_{1}^{\dagger}(g)AU_{1}(g)}=P_{a}^{U_{1}^{\dagger}(g)AU_{1}(g)}\rho P_{a}^{U_{1}^{\dagger}(g)AU_{1}(g)} \propto P_{a}^{U_{1}^{\dagger}(g)AU_{1}(g)}$, where, $P_{a}^{U_{1}^{\dagger}(g)AU_{1}(g)}$ is the projector of the observable $U_{1}^{\dagger}(g)AU_{1}(g)$ corresponding to outcome $a$. Again, $U_{2}(g)\tilde{\rho}_{a}^{A}U_{2}^{\dagger}(g)=U_{2}(g)P_{a}^{A}\rho P_{a}^{A}U_{2}^{\dagger}(g)\propto U_{2}(g)P_{a}^{A}U_{2}^{\dagger}(g)$. Now, two pure states $P_{a}^{U_{1}^{\dagger}(g)AU_{1}(g)}$ and $U_{2}(g)P_{a}^{A}U_{2}^{\dagger}(g)$ ar
 e always connected by some unitary, rather they are identical when $U_{2}(g)=U_{1}^{\dagger}(g)$. Hence, conditions of lemma 1 are satisfied.

With the help of this lemma we now show that under unitary evolution and projective measurements HSE and HVM in the context of temporal correlations are equivalent. To this end let us take the set of all pure states, $\lbrace |\lambda\rangle\in C^{d}| |\langle\lambda|\lambda\rangle|=1\rbrace$, for constructing HVM. The set of pure states together with the probability measure taken as the Haar measure over the unitary groups defines an unique optimal covariant ensemble.\\

{\bf Theorem 2}: \emph{For arbitrary initial state under unitary evolution and projective measurements HSM and HVM for the temporal correlations are equivalent}.\\

 {\bf Proof}: If there is a HSM for temporal correlation then it must also have a HVM, i.e. HSM $\Rightarrow$ HVM. This follows trivially since Eq.(\ref{ns}) resembles  Eq.(\ref{nirm})  by simply denoting $ p(b_{j}|B_{j},\lambda)= p^{Q}(b_{j}|B_{j},\rho_{\lambda})$ . Now we show for projective measurements and unitary evolution that the converse of the above implication, {\it i.e.,} HVM $\Rightarrow$ HSM is also true. To this end it is sufficient to show that if a HVM exists, then the states at Bob's hand also has a HSE, or in other words, Alice can simulate $\tilde{\rho}_{a}^{A}$ using the HSE, $\lbrace p_{\lambda}, \rho_{\lambda}\rbrace $, with the same $p_{\lambda}$ and $p(a|A,\lambda)$ appearing in the HVM.
 
 In the steering protocol defined in Sec. II, Bob asks Alice to measure $A$, and after measuring she announces the outcome $a$. Then Bob gets an unnormalised state $\tilde{\rho}^{A}_{a}=p(a|A)\rho^{A}_{a}$, where $\rho^{A}_{a}$ is the eigenstate of $A$ with eigenvalue $a$. Now, for unitary evolution and projective measurements as the lemma is satisfied for arbitrary initial states, there exists an optimal ensemble, and  without loss of generality let it consist of pure states with the Haar measure $\lbrace \rho_{\lambda}^{\star},p_{\lambda}^{\star}\rbrace$, s.t. $\tilde{\rho}_{a}^{A}=\sum p_{\lambda}^{\star}\rho_{\lambda}^{\star}p^{\star}(a|A,\lambda)$. From the existence of the HVM, we have $p(a|A)=\sum_{\lambda} p_{\lambda}p(a|A,\lambda)$, and since the optimal ensemble exists, we must have  $\sum_{\lambda} p_{\lambda}p(a|A,\lambda)=\sum_{\lambda} p_{\lambda}^{\star}p^{\star}(a|A,\lambda)$. As $\rho_{\lambda}$ and $\rho_{\lambda}^{\star}$ are pure states, they are unitarily
  related  and the
 invariance of the Haar measure over all spherical rotations implies Alice can construct another HSE, $\lbrace p_{\lambda},\rho_{\lambda}\rbrace$, with $p_{\lambda}=p_{\lambda}^{\star}$.  Consequently, we have $p(a|A,\lambda)=p^{\star}(a|A,\lambda)$. Thus from the knowledge of HVM Alice can simulate $\tilde{\rho_{a}^{A}}$. Hence, the theorem.   $\square$\\
 The above theorem states that under unitary evolution and projective measurements the existence of  HSM implies the existence of HVM, and {\it vice-versa}, from which it logically follows that violation of LGI implies violation of TSI,
 and {\it vice-versa}.
 
{\emph{An example with qubit}:} The above theorem is most general, implying that whatever be the dimension of the system, HVM and HSM are equivalent under unitary evolution and projective measurements. Now we present an example for a two level system showing that under unitary evolution and dichotomic projective measurements, LGI and TSI are both  violated under unitary evolution and projective measurements.
Note first, that it is possible to derive a
\emph{temporal CHSH inequality}~\citep{bv} when the two observers Alice (Bob) choose to measure between their corresponding  dichotomic observables $A_{1}$ and $A_{2}$ ($B_{1}$ and $B_{2}$) at times $t_{A}$ ($t_{B}$) respectively. It has been  shown under general conditions
that for dichotomic measurements the maximal value for LGI is obtained on qubits~\citep{bdrni}. Hence, it suffices to consider qubit systems as far as dichotomic observables and the maximal value of LGI are concerned. With these measurements 
the temporal CHSH inequality derived  in analogy with spatial one is given by~\citep{bv}
\begin{equation}\label{tchsh}
|E(A_{1},B_{1})+ E(A_{1},B_{2})+E(A_{2},B_{1})-E(A_{2},B_{2})| \leq 2,
\end{equation} 
 where the two time correlation functions are given by $E(A_{i},B_{j})=p(a_{i}=b_{j})-p(a_{i}\neq b_{j})$. Recently, a steering inequality for
spatial correlations has been derived~\citep{achsh} that is analogous to the CHSH inequality in the 
Bell-nonlocality scenario, in the sense that it forms a  necessary and sufficient condition for
the case of two parties and dichotomic measurements. The inequality for spatial correlations obtained
in Ref.~\citep{achsh} can be derived also in the scenario of temporal correlations involving single
system steering using the joint probability distribution (\ref{ns}) (in a similar way as the
joint probability distribution (\ref{nirm}) leads to a temporal analogue of the CHSH inequality or
the LGI). Thus, the following inequality holds when Alice prepares the state for Bob from a HS 
ensemble:
\begin{figure}[t!]
\centering
\includegraphics[height=5cm,width=8cm]{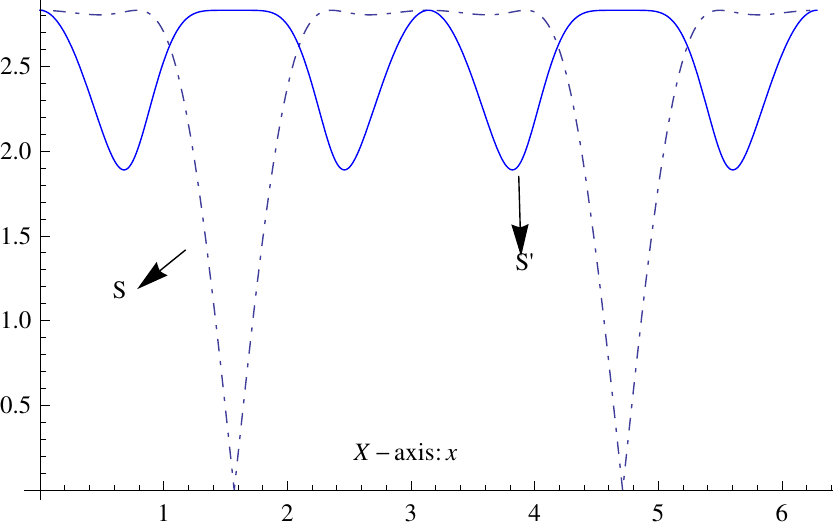}
\caption{(Color on-line) The steering sums $S$ are plotted versus the parameter $x=\omega\delta t$.  
The dot-dashed curve denotes $S$. The solid curve denotes $S'$. }\label{fig} 
\end{figure}
\begin{eqnarray}\label{nsi}
S=\sqrt{\langle (A_{1}+A_{2})B_{1}\rangle ^{2}+\langle (A_{1}+A_{2})B_{2}\rangle ^{2} }\nonumber\\
+ \sqrt{\langle (A_{1}-A_{2})B_{1}\rangle ^{2}+\langle (A_{1}-A_{2})B_{2}\rangle ^{2} }\leq 2
\end{eqnarray}
The usual LG scenario with unitary evolution of the form $U(t)=e^{-iHt/\hslash}$, and measurements of the same observable (say $Q$) at different times can be described as two sequential measurements in a row in the following way. For the choices of measurements $A_{1}=Q(t_{1}=\delta t), A_{2}=Q(t_{3}=3\delta t), B_{1}=Q(t_{2}=2\delta t), B_{2}=Q(t_{4}=4\delta t)$, 
where $Q(t)=\sigma_{z}\cos\omega t+\sigma_{x}\sin\omega t$, the steering function $S$ defined in 
Eq.(\ref{nsi})  is given by  $S = \sqrt{5\cos^{2}x+\cos^{2}3x+2\cos 3x\cos x}+\sqrt{\cos^{2}x+\cos^{2}3x-2\cos 3x\cos x}$. Here $x=\omega\delta t$, with $\delta t$ being the time difference between two 
successive measurements taken to be the same for all pairs of successive measurements. Now, permuting
the times  $t_{1}$ and $t_{4}$ in Eq.(\ref{nsi}), we get $S' =\sqrt{\cos^2 x+2\cos^2 2x+\cos^2 3x+2\cos 2x(\cos x+\cos 3x)}+\sqrt{\cos^2 x+2\cos^2 2x+\cos^2 3x-2\cos 2x(\cos x+\cos 3x)}$. In Fig.2 we plot the two 
steering sums $S$
and $ S'$ versus the parameter $x$. It can be seen from Fig.2 that at least one of the steering
sums is greater than the no-steering bound, and hence, the steering inequality (\ref{nsi}) is
violated for any value of $x$.

Now we show with the above choice of measurements, that an LGI is also violated for any choice of time interval between successive measurements.
  The inequality (\ref{tchsh}) is violated maximally for $A_{1}=\frac{B_{1}+B_{2}}{\sqrt{2}}$ and 
$A_{2}=\frac{B_{1}-B_{2}}{\sqrt{2}}$. 
These measurements can be mapped as, 
$Q(\omega t=0)=A_{1}=\sigma_{z}$;  $Q(\pi/4)=B_{1}=(\sigma_{z}+\sigma_{x})/\sqrt{2}$; 
$ Q(\pi/2)=A_{2}=\sigma_{x}$; and  $Q(3\pi/4)=B_{2}=(\sigma_{x}-\sigma_{z})/\sqrt{2}$. With permutation of 
the time indices there exist three  different $4$-term LGIs  and as shown in \citep{nori}, there is 
violation of at least any one of these inequalities for an arbitrary choice of the time interval of 
successive measurements. 
hence, under unitary evolution, for arbitrary time intervals `$\delta t$' of successive dichotomic sharp measurements, there is always violation of TSI and LGI.

%%%%%%%%%%%%%%%%%%%%%%%%%%%%%%%%%%%%%%%%%%%%%%%%%%%%%%%%%%%%%%%%%%%%%%%%%%%
%%%%%%%%%%%%%%%%%%%%%   up to here  %%%%%%%%%%%%%%%%%%%%%%%%%%%%%%%%%%%%
%%%%%%%%%%%%%%%%%%%%%%%%%%%%%%%%%%%%%%%%%%%%%%%%%%%%%%%%%%%%%%%%%%

\section{IV. Probing hierarchy of temporal correlations}

\subsection{A. Unitary evolution and generalised measurement} 

From the previous Section it is clear that by considering unitary evolution and projective measurement together hierarchy of temporal correlation cannot be probed. We now consider generalised measurements (POVMs) on the system undergoing unitary evolution in order 
to show that the two types of temporal correlations characterised by the violation of TSI and LGI respectively, are inequivalent in this scenario.

For the case of spatial correlations it has been shown~\citep{jntst} that 
if Alice's observables are jointly 
measurable, the joint statistics can be reproduced by a local model for any bipartite state and any 
measurement of Bob. Moreover, for any set of  POVMs at Alice's side that is not jointly measurable, there 
exists a  bipartite state and a set of measurements for Bob such that the resulting joint statistics 
violates a Bell inequality.
It is possible to consider joint measurement of a set of POVMs even when they do not commute. A set of POVMs ${M^{k}(a_{k})}$ is said to be compatible if 
their outcome statistics can be found as marginal of a global POVM $\lbrace G(\lambda); G(\lambda)\geq 0; \sum_{\lambda}G(\lambda)=1\rbrace $ statistics. Here $\lambda=(a_{1}, a_{2},...a_{n})$ and $M^{k}(a_{k})=\sum_{i\neq k}G(\lambda)$. From this global POVM marginal statistics can be obtained through classical post processing of 
grand statistics~\citep{ali}, $p(a_{k})=\sum_{\lambda}g(\lambda)p(a_{k}|\lambda)$, $g(\lambda)=tr[\rho G(\lambda)]$.
 For the case of two dichotomic POVMs non-joint measurement is necessary and 
sufficient for demonstrating steering and Bell nonlocality~\citep{wolf}. However, for three dichotomic POVMs 
not satisfying full but with pair wise joint measurability,  there is no violation of a large class of Bell type 
inequalities, whereas steering can be shown for this scenario~\citep{jntst}.

For temporal correlations with two dichotomic measurements it has been shown~\citep{sa,mal} that jointly 
measurable observables can not lead to LGI violation. Necessity and sufficiency of nonjoint measurability to 
demonstrate temporal steering has been demonstrated~\citep{udevi}. Here we find that there exist nonjoint 
measurable observables that can demonstrate steering without leading to LGI violation.

 A quadratic steering inequality for measurements in $N=2$ or $3$ mutually unbiased basis is 
given by~\citep{tepster}
\begin{eqnarray}\label{qstr}
S_N=\sum_{i=1}^NE[\langle B_{i}\rangle^{2} _{A_{i}}]\leq 1.
\end{eqnarray}
where, 
$E[\langle B_{i}\rangle^{2} _{A_{i}}]=\sum_{a_{i}=\pm 1}p(A_{i}=a_{i})\langle B_{i}\rangle _{A_{i}=a_{i}}^{2}$,
with
$p(A_{i}=a_{i})$ being the probability of getting $a_{i}$ at $t_{A}$, and $\langle B_{i}\rangle _{A_{i}=a_{i}}^{2}$ 
is the expectation value of $B_{i}$ at $t_{B}$ on the state measured by Alice at $t_{A}$.
Let us consider three dichotomic POVMs acting on the two dimensional Hilbert space as 
$M^{k}(a_{k})=\frac{1}{2}(\mathbb{I}+\eta a_{k}\sigma_{k})$. This is an example of an unsharp measurement with 
sharpness parameter $\eta$~\citep{bush}, where $k\in \lbrace x, y,z\rbrace $, and $\sigma_{k}$ are Pauli 
matrices.
We make the following choice of the observables: $A_{1}=\eta\sigma_{z}, A_{2}=\eta\sigma_{y}$, and 
$A_{3}=\eta\sigma_{x}$. The system evolves under the Hamiltonian $U=e^{-i\sigma_{x}\omega t/2}$ when $A_{1},A_{2}$ are measured and $V=e^{-i\sigma_{y}\omega t/2}$ when $A_{3}$ is measured. Going to the Heisenberg picture, Bob's
observables are given by $B_{1(2)}=U^{\dagger}\sigma_{z}(\sigma_{y})U$, and $B_{3}=V^{\dagger}\sigma_{x}V$.
With the above choices we get $S_{3}=3\eta^{2}\cos^{2}\theta$.  It is now straightforward to see that 
Eq.(\ref{qstr}) is violated for $\eta > \frac{1}{\sqrt{3}}(=0.57735)$.

We now consider a class of LGI  is given by (see Ref.~\citep{nori} and references therein)
\begin{eqnarray}\label{glgi}
K_{n}=C_{21}+C_{32}+...+C_{n(n-1)}-C_{n1}
\end{eqnarray}
where, $K_{n}$ is the $n$-term LG sum derived from outcome statistics of measurements of an observable, $Q$ at
 times $t_{1}, t_{2}...t_{n}$, and $C_{ij}$ is the correlation between two sequential measurements. Under the
assumptions of macrorealism this quantity is bounded by
$-n\leq K_{n}\leq n-2; \>\> n\geq 3$, for odd $n$, and by 
$-(n-2)\leq K_{n}\leq n-2; \>\> n\geq 4$, for even $n$.
It has been shown~\citep{mal} that for $\eta\leq \sqrt{(n-2)/(n\cos\frac{\pi}{n}})$,  no 
violation can be found.

As we want to compare with the three measurement steering scenario, the relevant LGIs are $K_{5}$ and $K_{6}$. 
Both of them can be mapped to a situation where Alice and Bob measure three different observables on the same system at time $t_{A}$ and $t_{B}$ sequentially,
with no time evolution of the state between the measurements of Alice and Bob. 
We consider the LGI
\begin{equation}\label{k5}
K_{5}=C_{21}+C_{32}+C_{43}+C_{54}-C_{51}.
\end{equation}
In fact, without loss of generality any higher order LG test on a single system can thus be mapped to a series of two sequential measurements where Alice chooses to measure first and then Bob from different sets of observables. This mapping  scenario is different from the one discussed in \citep{fritz}, and in the latter the mapping from spatial correlation to temporal correlation  cannot be extended beyond  more than two outcomes and two parties. For sequential measurements one can show that for macrorealist and noncontextual theories the bound is the same, i.e., $K_{5}\leq 3$ which can be violated in quantum mechanics. It is shown in \citep{bdrni} that a Tsirelson like bound for macrorealist theory is $4.04$, whereas for noncontextual theories it is $3.94$.

 Here, in order to reproduce two point correlations in eq.(\ref{k5}) yielding maximal violation in the mapped situation, Alice's choice of measurements are the three observables $A_{1}=Q(\pi/5), A_{2}=Q(3\pi/5), A_{3}=Q(\pi)$, and 
Bob's choices are  $B_{1}=Q(2\pi/5), B_{2}=Q(4\pi/5), B_{3}=Q(\pi/5)$,  where $Q(\theta)=(\sigma_{z}\cos{\theta}+\sigma_{x}\sin{\theta})$. 
The correlation $C_{21}=\langle A_{1}B_{1}\rangle$ means Alice first measures $A_{1}$ and then Bob measures $B_{1}$ sequentially on the same system. Similarly for $C_{32}=\langle A_{2}B_{1}\rangle, C_{43}=\langle A_{2}B_{2}\rangle, C_{54}=\langle A_{3}B_{2}\rangle, C_{51}=\langle A_{3}B_{3}\rangle$. For this kind of mapping to  hold it is crucial that for dichotomic observables, $C_{ij}$s are independent of the order of the measurements as, $C_{ij}=\frac{1}{2}tr[\rho\lbrace A_{i},B_{j}\rbrace]$ \citep{fritz}.  

 It turns out that for $\eta \leq 0.861186$, no  violation of the LGI is possible in this case. 
From the previous discussion it is clear that temporal steering is possible for $\eta > 0.57735$ as 
$S_{3} > 1$. Hence, in the range $0.861186>\eta>0.57735$ steering can be shown but no LGI violation can 
be demonstrated. If we consider $K_{6}=C_{21}+C_{32}+C_{43}+C_{54}+C_{56}-C_{61}$, Alice's choices are $A_{1}=Q(\pi/6), A_{2}=Q(\pi/2), A_{3}=Q(5\pi/6)$ and Bob's are $B_{1}=Q(\pi/3), B_{2}=Q(2\pi/3), B_{3}=Q(\pi/6)$.
In this case in the range $0.877383>\eta>0.57735$ steering can be demonstrated but not LGI violation. In the context of three measurement settings it is sufficient to consider $K_{5}$ and $K_{6}$. Even if we consider all the higher order LGIs, it is straightforward to see from the upper bound of $\eta$ given above, that the range  for which steering can be shown but no LGI violation can be demonstrated  increases, thus strengthening the inequivalence between them.

The hierarchy between the two types of temporal correlations can be generalized to higher dimensional 
systems as well. To demonstrate temporal steering non-joint POVMs are required, whatever be the 
dimension of the system or the cardinality of the outcome set of measurements~\citep{udevi}. Temporal 
steering witness for higher dimensions with nondegenerate measurements has  been derived 
recently~\citep{stwit}.  It is further possible to show that there exist non-joint measurements 
for which  
LGI is not violated under the restriction of dichotomic measurements. For spin $j$ systems the parity 
operator as the dichotomic observable was considered in Ref.~\citep{kb}, and the maximum value of 
the four term LGI was obtained asymptotically to be  $2.481$. In Ref.~\citep{sa} this bound is improved 
to $2\sqrt{2}$ which is optimal for dichotomic measurements in any dimension by considering a different 
measurement scheme. In such a  scheme~\citep{sa} the multilevel spin $j$ system is transformed into 
$(2j+1(2))/2$  two level systems for $2j+1$ even (odd). Then each system is evolved separately and 
measured subsequently by the application of operators acting on two dimensional Hilbert spaces. The
relevant observable is given by $Q=\frac{\Gamma_{z}  + \Pi}{\sqrt{2j+1}} $, where $\Pi$ is a null 
matrix when $2j+1$ is even, and for odd  $2j+1$, the only nonvanishing element of $\Pi$ is 
$(\Pi)_{N,N}=\frac{1}{\sqrt{2}}$. Here $\Gamma_{z}$ is block diagonal matrix with $\sigma_{z}$. Time 
evolution of the separated two level systems are affected by $U(t)=\exp^{-i\omega t\sigma_{x}/2}\oplus \exp^{-i\omega t\sigma_{x}/2}\oplus ...\oplus \exp^{-i\omega t\sigma_{x}/2}$. In such a scenario  all the treatment of 
two level systems described above in the present work follows.

\subsection{B. Sharp measurement under nonunitary evolution}

\begin{figure}[t!]
\centering
\includegraphics[height=5cm,width=8cm]{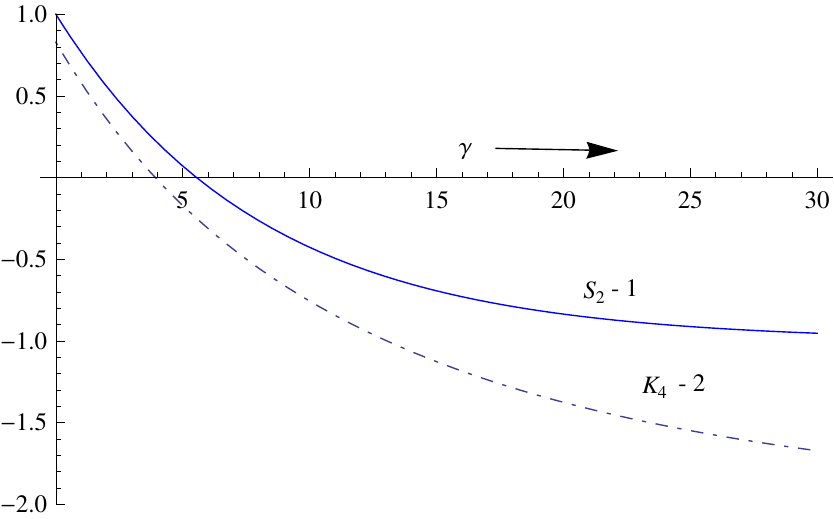}
\caption{(Color on-line) The LG sum $K_{4}-2$ (dotdashed curve)  and
the temporal steering sum $S_{2}-1$ (solid curve) are plotted respectively versus the damping parameter
$\gamma$ . Positive value of these curves indicates violation of the respective inequalities. 
Here $\omega\delta t=\pi/4$.}
\end{figure}

We now show that with sharp measurement under noisy evolution,  temporal steering is possible even when 
the violation of LGI is washed out by noise. Consider a qubit undergoing Rabi oscillation sent through an amplitude damping channel to Bob. The Markovian decay process in Lindblad form is described by 
\begin{eqnarray}
\frac{d\rho}{dt}=-\frac{i}{\hslash}[H,\rho]+\frac{\gamma}{2}(2\sigma_{-}\rho\sigma_{+}-\sigma_{+}\sigma_{-}\rho+ \rho\sigma_{+}\sigma_{-})
\end{eqnarray}
where $\gamma$ is the noise parameter. Taking $H=-\frac{\omega}{2}\sigma_{z}$ and $\hslash=1$ with a 
maximally mixed initial state, the two time correlation for measurement of $\sigma_{x}$ is given by 
$\cos(\omega\delta t)\exp^{-\gamma\delta t}$. The corresponding four term LG sum ($K_{4}$, see eq.(\ref{glgi})) becomes
\begin{eqnarray}
K_{4}=3 \exp^{-\gamma\delta t}\cos(\omega\delta t) - \exp^{-3\gamma\delta t}\cos(3\omega\delta t )\leq 2.
\end{eqnarray}
For Alice and Bob's choice of measurement given by
 $A_{1}=\eta\sigma_{z}, A_{2}=\eta\sigma_{y}$,  $B_{1}=U^{\dagger}\sigma_{z}U$,
$B_{2}=U^{\dagger}\sigma_{y}U$,
 the
steering parameter $S_{2}$ given by eq.(\ref{qstr}), becomes
\begin{eqnarray}
S_{2}=2\exp^{-2\gamma\delta t}\cos^{2}(\omega\delta t)\leq 1.
\end{eqnarray}
In Fig.~3 we plot the functions $K_4-2$ and $S_2-1$ versus the damping parameter $\gamma$. 
It is clear from the figure that after the damping parameter $\gamma$ exceeds a certain value, the
violation of LGI disappears, but temporal steering persists upto a greater value of $\gamma$.\\

\section{V. Conclusions} 

In this work we have performed a comparative study of two different kinds of temporal correlations
in quantum mechanics: (i) those responsible
for single system steering or temporal steering, and (ii) those responsible for the violation
of Leggett-Garg inequalities. 
Any hidden state model gives rise to a hidden variable model, and hence, whenever there is LGI violation there would be TSI violation. The nontrivial result which we obtain here is that when the system evolves under unitary operations and only projective measurements are
allowed, the reverse is also true {\it i.e.}, TSI and LGI violating correlations are equivalent.

 Next, in
order to exhibit the hierarchy between the two types of correlations, we consider again two
separate scenarios allowing for either unsharp measurements (POVMs) or noisy evolution.
We find in the former case that when three dichotomic measurements are performed by the observers there exist non-joint measurements for which steering can be 
demonstrated whereas no violation of LGIs pertinent to the scenario can be shown. This feature is generalized to arbitrary 
dimensional systems under the restriction of dichotomic measurements. However, it suffices to consider two level system as far as dichotomic observables are concerned. In the case of non-unitary
evolution with projective measurements, we show that temporal steering is more robust against noise compared to the violation of LGI,
again indicating the hierarchy between the two types of correlations. 

Before concluding, we note that
the hierarchy revealed here for temporal correlations is somewhat analogous to the hierarchy of spatial 
correlations, but with certain key differences. In the case of spatial correlations the hierarchy
between steering and Bell-nonlocality is hidden at the level of pure states, but revealed through
the use of mixed states. On the other
hand, in the context of temporal correlations, the hierarchy is not at the level of states. Here the
hierarchy at the level of correlation between sequential measurements, is hidden if one considers both
unitary evolution and sharp measurements, and revealed if either of them is relaxed. It remains for future studies to explore how this hierarchy would
fare in the context of multiple outcome measurements. Finally, it may be interesting to formulate  protocols for information processing with differential degrees of security
or device-independence based on such a hierarchy. 

{\it\textbf{Acknowledgements:}} ASM and  DH acknowledge support from the project SR/S2/LOP-08/2013 of DST,
India.

\end{document}